\documentclass[10pt,pra,aps,showpacs,twocolumn,unsortedaddress,longbibliography]{revtex4-1}

\usepackage{graphicx,bm}                                                       
\usepackage{amsmath}                                                           
\usepackage{amssymb}                                                           
\usepackage{epsfig}                                                            
\usepackage{epsf}                                                              
\usepackage{verbatim}                                                          
\usepackage[usenames]{color}                                                   
\usepackage[colorlinks=true,linkcolor=blue]{hyperref}

\newcommand{\bi}{\bm}

\def\be#1\ee{\begin{equation}#1\end{equation}}
\newcommand{\ba}{\begin{eqnarray}}
\newcommand{\ea}{\end{eqnarray}}
\newcommand{\dd}{\mathrm{d}}
\newcommand{\bvk}{\bi{k}}

\begin{document}
\title{Superfluidity and spin superfluidity in spinor Bose gases}

\author{J. Armaitis}
\email{jogundas.armaitis@tfai.vu.lt}                                           
\affiliation{Institute of Theoretical Physics and Astronomy,
Vilnius University,
Saul\.{e}tekio Ave.\ 3, LT-10222 Vilnius, Lithuania}

\author{R. A. Duine}
\affiliation{Institute for Theoretical Physics and Center for Extreme
Matter and Emergent Phenomena,
Utrecht University,
Princetonplein 5, 3584 CC Utrecht, The Netherlands}
\affiliation{Department of Applied Physics, Eindhoven University of Technology, P.O. Box 513, 5600 MB Eindhoven, The Netherlands}

\begin{abstract}
We show that spinor Bose gases subject to a quadratic Zeeman effect exhibit coexisting superfluidity and spin superfluidity, and study the interplay between these two distinct types of superfluidity. 
To illustrate that the basic principles governing these two types of superfluidity
are the same, we describe the magnetization and particle-density
dynamics in a single hydrodynamic framework. In
this description spin and mass supercurrents are
driven by their respective chemical potential gradients.
As an application, we propose an
experimentally-accessible stationary state, where the two types of supercurrents counterflow and cancel each other, thus resulting
in no mass transport. Furthermore, we propose a straightforward setup to probe spin superfluidity
by measuring the in-plane magnetization angle of the whole cloud of atoms.
We verify the robustness of these findings by
evaluating the four-magnon collision time, and find that the timescale
for coherent (superfluid) dynamics is separated from that of the 
slower incoherent dynamics by one order of magnitude.
Comparing the atom and magnon kinetics reveals that while the former
can be hydrodynamic, the latter is typically collisionless under most 
experimental conditions.
This implies that, while our zero-temperature hydrodynamic equations are a valid description of spin transport in Bose gases, a hydrodynamic description that treats both mass and spin transport at finite temperatures may not be readily feasible. 
\end{abstract}

\maketitle

\section{Introduction} 
The phenomenon of superfluidity underlies transport 
properties of numerous systems, including various superconductors
\cite{Bennemann13}, liquid helium \cite{KhalatBook}, both bosonic 
\cite{PethickSmithBook} and fermionic \cite{Fermi2ndSound13}
ultracold atoms, exciton-polariton condensates 
\cite{Deng10}, topological insulators \cite{Qi2009,Tilahun2011,Peotta2015}, as 
well as neutron stars \cite{Dean2003} and flocks of birds 
\cite{Attanasi2014}. The possibility to achieve dissipationless 
propagation of information at room temperature has recently fueled
interest in spin superfluidity \cite{Sonin2010, Takei2014} in general and
in magnon spintronics \cite{Duine2015,Cornelissen2015,Chumak2015}
in particular.

Ferromagnetic spinor Bose-Einstein condensates of atomic vapour
stand out among these systems as a rare example where two types of 
superfluidity can be present simultaneously, and where they are also readily 
experimentally addressable.
Specifically, experimental ultracold atom techniques currently allow
controlled excitation and imaging of both the local phase
of the condensate, and also of the spin texture. 
Exciting mass supercurrents (pertaining to inhomogeneity of the local phase of the wavefunction)
in the system is possible by, e.g., stirring the condensate with an ``optical spoon'' \cite{Madison2000}.
Signatures of this mass superfluidity have been observed
in the collective mode spectrum \cite{Marago2000}
and lattices of quantized vortices \cite{Abo-Shaeer2001}.
Furthermore, manipulating and observing the spin texture
(or, equivalently, the spin supercurrent \cite{Sonin2010})
has recently also become possible in this system. In particular,
spin-agnostic optical traps \cite{Stamper-Kurn1998}
have allowed preparation and subsequent imaging
of spinor gases, for example, using
the Stern-Gerlach method \cite{Stenger1998}
and also directly \cite{Sadler2006}.
Several methods for imprinting spin textures have been developed, either
relying on varying external magnetic fields \cite{Choi2013} or optical transitions \cite{CoherentMagnonOptics}.

\begin{figure}[ht]
\begin{center}
\includegraphics[width=0.9\linewidth]{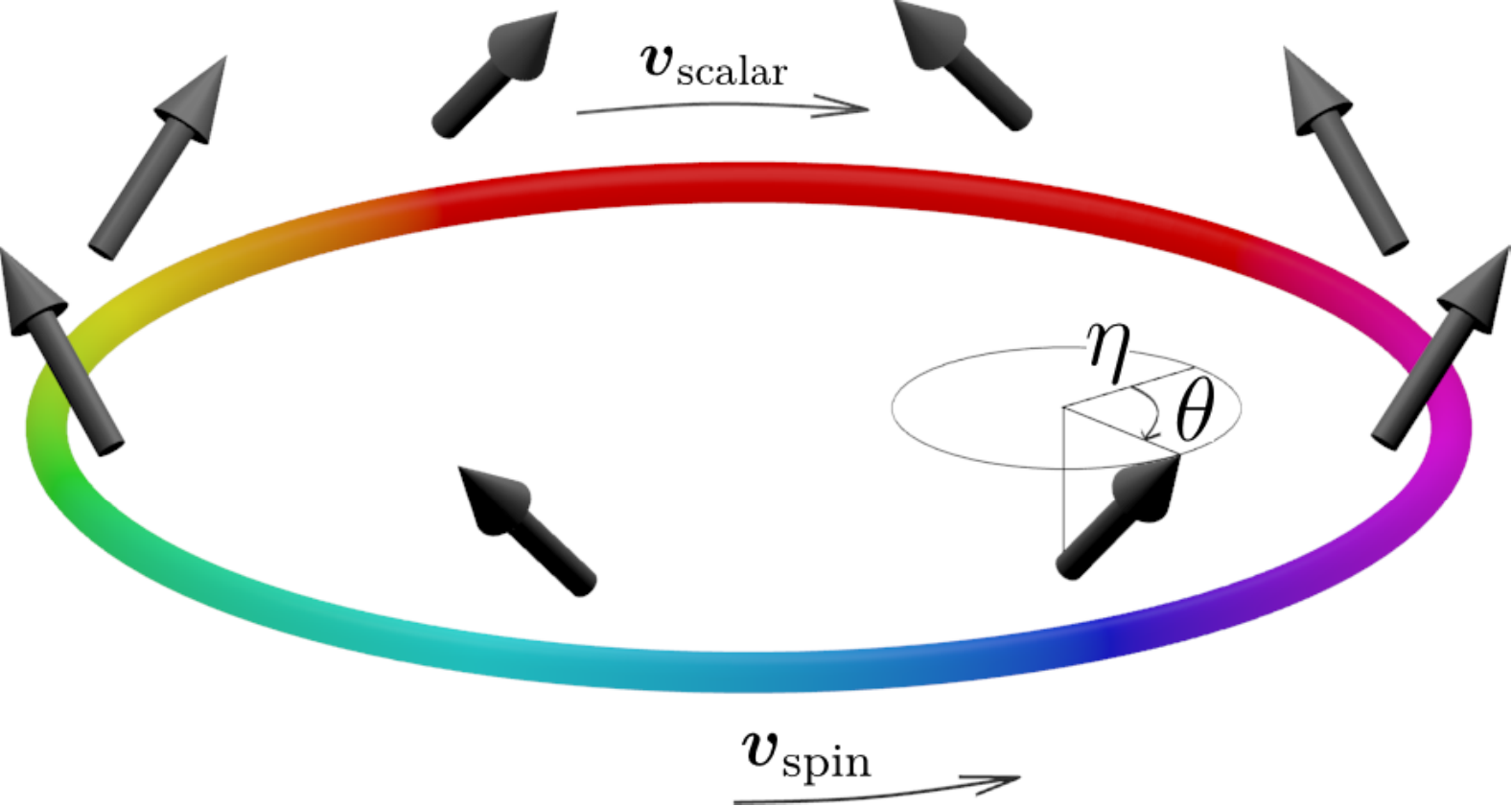}
\caption{
By counterflowing, a supercurrent and a spin supercurrent create a stationary
state with no mass flow in a spinor Bose gas on a ring. In a uniform-density
system, the supercurrent $\bi v_\mathrm{scalar}$ is due to the global phase
texture of the atomic condensate, whereas the spin
supercurrent $\bi v_\mathrm{spin}$ is due to the in-plane magnetization texture
(indicated by arrows which can be recast in terms of a phase $\theta$) of the
magnon condensate with a measure of the condensate fraction $\eta$.  The two
textures can be independently engineered in a ferromagnetic spinor Bose gas.
}
\label{fig:exp0}
\end{center}
\end{figure}

Earlier work \cite{Cherng2008} has investigated stability of the planar spin spiral (XY spiral), which is one of the 
states that we consider (Fig.~\ref{fig:exp0}). Both adiabatic and sudden preparation was carefully considered, and it was 
demonstrated that this state is stable for sufficiently small spiral wave vectors. However, despite the recent observation of a
(quasi-)condensate of magnons \cite{Yukalov2012,Fang2015}, 
few studies have been devoted to spin superfluidity in ultracold spinor gas 
\cite{Ashhab2005,Kudo2011,Flayac2013,Zhu2015},
even though this phenomenon has played a prominent role in liquid helium systems \cite{Bunkov2010}. 
Furthermore, the interplay of mass and spin superfluidity
has not been addressed in the ultracold-atom context to the best of our knowledge. 

In this article we examine coherent dynamics of spinor Bose gases, show that these systems may exhibit coexisting superfluidity and spin superfluidity, and study their interplay.
We construct a hydrodynamic model that incorporates both types of superfluidity on equal footing. In linear response, they are decoupled. We determine both the collective mode spectrum and demonstrate that spin superfluidity can be experimentally observed by monitoring the in-plane  magnetization angle of the atoms. Moreover, we demonstrate that non-linear effects lead to experimentally stationary states where both supercurrents are counterflowing.
Finally, we check the robustness of our
findings by comparing the timescales relevant to these coherent processes with the collision
timescales describing incoherent dynamics of this system
and show that they are well separated.

\section{Theoretical framework} 
\label{sec:framework}
We consider a spinor
Bose gas described by the second-quantized Hamiltonian
\cite{UedaHydro,Stamper-Kurn2013,Ueda2012}
$\hat H = \hat H_0 + \hat H_Z + \hat H_I$, where
$
\hat H_0 = \int d\bi x \,
\hat \psi_\mu^\dagger
\left(
-{\hbar^2\bi \nabla^2}/{2M} + V(\bi x)
\right)
\hat \psi_\mu
$
represents the kinetic energy and the trapping potential $V(\bi x)$. The field operators $\hat \psi_\mu$ correspond 
to atoms with mass $M$ in the hyperfine state $\mu$. (We use the Einstein summation convention throughout the 
article.) The linear and quadratic Zeeman effects are described by
$
\hat H_Z = -\int d\bi x \,
\hat \psi_\mu^\dagger
\left(
p [\sigma_z]_{\mu\nu} - \hbar K [\sigma_z^2]_{\mu\nu}
\right)
\hat \psi_\nu,
$
where $p$ ($\hbar K$) is the energy of the linear (quadratic) Zeeman effect%
\footnote{                                                                     
We concentrate on the regime where $p>0$ and $K>0$ in this paper,
see also Sec.~\ref{ss:magcon}.
}, and
$\boldsymbol \sigma_{\mu\nu}$ is a vector of spin matrices. 
The external magnetic field has been chosen to point in the $z$-direction.
We only consider the ferromagnetic state, i.e., the situation where the energy 
of the linear Zeeman effect is larger than the one of the quadratic Zeeman effect 
($p>K$) \cite{Murata2007, Phuc11}.
Our present treatment is confined to the spin-1 case, which is the lowest-spin
system where the quadratic Zeeman term is nontrivial. Extension of our results
to higher-spin systems under the same approximations is straightforward.
The interactions between particles are described by
$
\hat H_I = \int d\bi x \,
\left(
g_0 
\colon\hat \rho^2\colon
\,
+
\,
g_1 \colon(\hat {\bi n}\hat {\rho})^2\colon
\right)/2,
$
where the colons denote normal ordering, 
$g_0$ is the spin-independent interaction strength,
$g_1$ is the spin-dependent interaction strength,
$\hat \rho = \hat \psi_\mu^\dagger \hat \psi_\mu$ is the density operator,
and 
$\hat {\bi n} = \hat \psi_\mu^\dagger {\boldsymbol \sigma}_{\mu\nu} \hat \psi_\nu / \hat \rho$ 
is the local spin operator. 
For completeness, we note that
the mean-field dynamics derived from this Hamiltonian
for the fields $\psi_\mu$ corresponding to the aforementioned field
operators $\hat \psi_\mu$ are
described by the Gross-Pitaevskii equation \cite{Ohmi1998,Ho1998},
\ba
i\hbar \partial_t \psi_\mu 
&=& 
\left(
-\frac{\hbar^2}{2M} \bi \nabla^2 + V(\bi x)
\right) \psi_\mu
\nonumber \\
&+&
\left(
-p [\sigma_z]_{\mu\nu} + \hbar K [\sigma_z^2]_{\mu\nu}
+g_0 \rho 
\right) \psi_\mu
\nonumber \\
&+&
g_1 \bi n \cdot {\boldsymbol \sigma}_{\mu\nu} \rho \psi_\nu,
\ea
where $\bi n$ is the mean-field magnetization, which corresponds to the operator
$\boldsymbol{\hat n}$ and is defined as ${\bi n} = \langle\hat \psi_\mu^\dagger {\boldsymbol 
\sigma}_{\mu\nu} \hat \psi_\nu / \hat \rho\rangle$.
The mean-field density is defined in a similar manner, $\rho = \langle \hat \rho \rangle$.
Zero-temperature hydrodynamic equations below are obtained
from this Gross-Pitaevskii equation.
We note at the outset that the quadratic Zeeman effect will turn out to be crucial in stabilizing the spin superfluidity.

The equations for $\langle\hat\psi_\mu\rangle$ governing the mean-field dynamics of this system at zero temperature can
be derived from the Hamiltonian $\hat H$ and written down in terms of a set of slowly evolving variables. 
At this point we restrict ourselves
to the ferromagnetic state with the saturated local magnetization $\bi n^2 = 1$, which implies
that we do not consider the nematic \cite{UedaHydro,Stamper-Kurn2013,Ueda2012} or antiferromagnetic \cite{Oh2014} evolution.
This approximation is well justified only when the gap introduced by the
spin-dependent interaction $\Delta_\mathrm{FM} = g_1 \rho$ is much larger
than all the other relevant energy scales,
in particular the exchange energy $\hbar J\pi^2/L^2$, 
easy-plane anisotropy energy $\hbar K$,
and the energy corresponding to the incoherent dynamics (see
Sec.~\ref{ss:incoherent}).
In that case, the dynamics are confined to the ferromagnetic manifold over the pertinent
time scales. 
Furthermore, we do not consider the trapping potential and the quantum pressure term \cite{PethickSmithBook}.
Moreover, we omit the linear Zeeman effect, as it can be removed by 
going to a rotating coordinate system.
In order to write down these mean-field equations in a concise manner,
we define the usual Eulerian derivative 
$D_t = \partial_t + \bi v \cdot \bi\nabla$, 
where the velocity $\bi v=-i(\hbar/2M)(\psi^*_\mu\bi \nabla \psi_\mu - [\bi \nabla \psi^*_\mu] \psi_\mu)/\rho$ governs mass transport but has 
contributions from both the global phase of the wavefunction 
and the spin texture. In particular, 
we find that magnetization dynamics in a ferromagnetic spinor Bose gas 
is described by a 
Landau-Lifshitz (LL) equation \cite{Lamacraft2008,Kudo2011}
\ba
D_t \bi n =& J \bi n \times \bi \nabla^2 \bi n
-
K \bi n \times \bi e_z n_z
\nonumber \\
&+ 
J (\bi n\times \nabla_i \bi n)(\nabla_i \rho)/\rho
\label{eq:lle}
,
\ea
where $\rho=\langle \hat \rho \rangle$ is the average density
(at zero temperature equal to the atomic condensate density)
and 
the exchange constant $J$ comes from the kinetic term 
in $\hat H$ and describes spin stiffness.
Neglecting interactions between spin waves, at low temperature
the spin stiffness is \cite{Kunimi2015}
$J=\hbar/2M$.
The magnetic anisotropy comes from the quadratic Zeeman effect. It can be generated using a sufficiently 
strong external magnetic field, in addition to radio-frequency and optical means \cite{Stamper-Kurn2013}. Three (spin-1) or more hyperfine states are required for the
magnetic anisotropy to be available in an atomic system. When $K<0$, it is favorable for the spin to align with the 
director of the magnetic field. In this case $z$ is known as the easy axis. 
On the other hand, in the so-called easy-plane situation $K>0$, the configuration with $\bi n$ perpendicular to 
the $z$ axis is energetically favored. Both of these situations can be achieved in a system of ultracold atoms \cite{Stamper-Kurn2013}.
In addition to the terms already present in the LL equation, various spin-relaxation terms may be added, such
as the transverse spin diffusion \cite{Armaitis2013} and Gilbert damping \cite{Gilbert2004}, 
as well as terms due to magnetic field inhomogeneities
\cite{ThermometryDemagCooling} and magnetic dipole-dipole interactions
\cite{Kawaguchi2007}. However, all these terms can be made small in an ultracold-atom system,
hence, we do not consider them here.

\section{Results}
\subsection{Magnon condensate}
\label{ss:magcon}                                                                 

\begin{figure}[t]
\begin{center}
\includegraphics[width=0.6\linewidth]{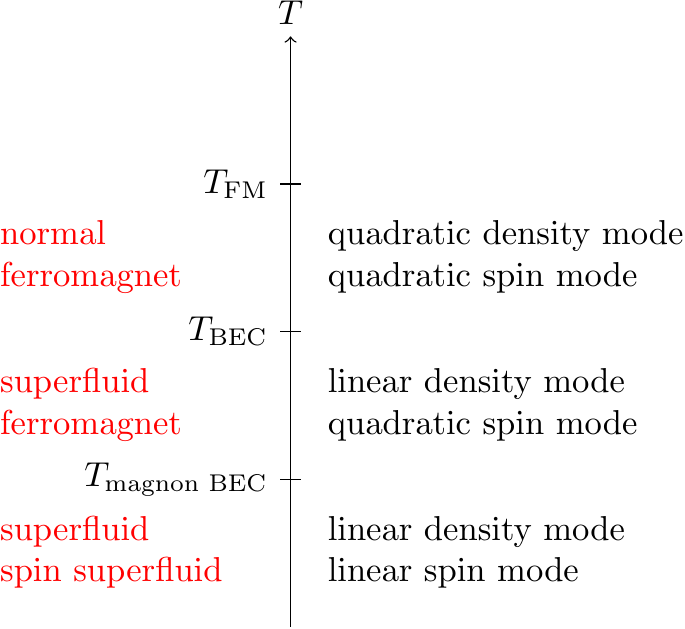}
\caption{
In a ferromagnetic spinor Bose gas in the presence
of a quadratic Zeeman effect, three phase transitions may occur as the temperature
is lowered. At high temperature the gas is thermal and has no spin order. 
Below $T_\mathrm{FM}$ the system enters the ferromagnetic phase, characterized
by the ordered magnetization (normal ferromagnet). 
However, both spin and density modes remain quadratic.
Below the condensation temperature $T_\mathrm{BEC}$ the Bose-Einstein condensate
forms, and one of the collective excitations gains a linear Bogoliubov dispersion 
(superfluid ferromagnet).
Finally, below the magnon-condensation temperature 
$T_\mathrm{magnon BEC}$, the dispersion
of the spin excitations also become linear (spin superfluid). 
The presence of two linear collective modes
in this low-temperature phase heralds the existence of two types of superfluid currents.
}
\label{fig:temp-col}
\end{center}
\end{figure}

In spinor Bose gases magnetization in the direction of the magnetic 
field ($n_z$ in this case) is conserved. In this paper, we consider 
bosons deep in ferromagnetic regime, but with magnetization not 
saturated in the direction of the magnetic field 
($n_z < |\boldsymbol n|$), 
and thus rotationally invariant around the $z$ axis for sufficiently 
high temperature. In our case, the phase diagram of the system is 
altered as compared to the usual treatment \cite{Murata2007}, 
where given the 
same magnetic field strength $p>K$ only the ferromagnetic phase is 
allowed. In particular, our setup implies that below a certain 
temperature $T_\mathrm{magnon BEC}$ rotational invariance 
around the $z$ axis 
is broken, precession of the magnetization around the magnetic field 
direction becomes synchronized, and the system is said to be in the 
magnon condensate phase. The schematic phase diagram in our case is 
given in Fig.~\ref{fig:temp-col}.

Deep in the ferromagnetic regime where the spin excitations are small
deviations from the average direction of the magnetization, it is natural
to describe the magnetization dynamics in terms of magnons.
Going to this description requires performing 
the Holstein-Primakoff transformation \cite{Holstein1940,auerbachbook}, 
which introduces bosonic magnon operators
(c.f.~Sec.~\ref{ss:incoherent}). This gas of magnons
can have a thermal component, and also a condensate component.
The condensate of magnons \cite{Demokritov2006,Troncoso2014} 
corresponds to the coherent magnetization
precession of the spins in the whole sample.
The symmetry that is broken when the magnon condensate forms concerns 
the in-plane magnetization angle $\theta$, and the absolute
value of the order parameter is related to the out-of-plane magnetization 
component $n_z$, cf.~Eq.~\eqref{eq:magBEC} and below.

Since we concentrate on the superfluidity
in this system, we do not describe the thermal magnons, and we also do
not investigate the complicated magnon condensation process \cite{Fang2015}.
However, we remark for completeness that in a ferromagnetic spinor Bose gas in the presence
of a quadratic Zeeman effect with $n_z$ conserved,
three phase transitions may occur as the temperature
is lowered (see Fig.~\ref{fig:temp-col}). The well-known Bose-Einstein condensation
in the mean-field theory occurs at \cite{PethickSmithBook}
\be
T_\mathrm{BEC} = \frac{2 \pi \hbar^2}{k_B M} \left(\frac{\rho}{\zeta_{3/2}}\right)^{2/3},
\ee
where $k_B$ is the Boltzmann constant and $\zeta_{3/2} \simeq 2.612$. 
The order parameter of the BEC phase is proportional to the
expectation value of the atom annihilation operator.

At a temperature $T_\mathrm{FM}$, which at the mean-field
level is higher than but similar to $T_\mathrm{BEC}$ \cite{Gu04,Gu03},
the system becomes ferromagnetically ordered. 
The global magnetization is the order parameter defining the ferromagnetic
state, and it becomes nonzero as the system enters the ferromagnetic state.
However, in the system we consider, $n_z$ is conserved, and thus the global
magnetization in the $z$ direction cannot be a function of temperature.
Therefore, the ferromagnetic transition manifests itself in a subtler than
usual manner. For example, spatial regions emerge with either a nonzero
in-plane magnetization \cite{Stenger1998}, or with different values of $n_z$
\cite{Hall1998}, see Ref.~\cite{Stamper-Kurn2013} for a more thorough
discussion. Most importantly for our purposes, deep in the ferromagnetic state
the local magnetization is fully saturated ($|\bi n|=1$) throughout the system
irrespective of conservation of $n_z$.

Finally, if $p<\hbar K$, the system enters the
magnon condensate 
phase at $T_\mathrm{magnon BEC}\simeq T_\mathrm{BEC}(1-p/\hbar K)^{2/3}$ \cite{Flebus2016}.
The magnon condensate is equivalent to the broken axisymmetry phase
in the full thermodynamic phase diagram of the system 
\cite{Murata2007,Phuc11}, with the crucial difference that
whereas the total magnetization is still saturated 
($|\boldsymbol n|=1$) in the magnon condensate phase,
in the broken axisymmetry phase it is not ($|\boldsymbol n|<1$).

In general, during the process of condensate formation, a certain symmetry
is broken, leading to a nonzero order parameter. In the case of Bose-Einstein condensation of atoms,
the symmetry concerns the phase of the wavefunction, and the order parameter is the average of the
atom-annihilation operator. In the condensate of magnons \cite{Demokritov2006,Troncoso2014}, the symmetry concerns
the in-plane magnetization angle $\theta$, and the order parameter is related to the out-of-plane magnetization component $n_z$.
It is important to
point out that the Landau-Lifshitz equation explicitly preserves the local magnetization 
$|\bi n|$ in stark contrast
to the dynamics of a BEC of atoms, where no such order-parameter conservation law exists in the grand-canonical
description \cite{LittleWhiteBook}.
Therefore, we expect the magnon condensate formation to differ
from the process of Bose-Einstein condensation of atoms.

In order to avoid various complications associated to the magnon condensate
formation, we propose to start from a sample that is 
homogeneously magnetized in the $z$ direction, and then
coherently tilt the magnetization (e.g., by an RF pulse) towards the
$x-y$ plane. This populates the pure-magnon-condensate state directly by 
preparing the magnetization of the system in the
following configuration:
\be
\bi n = (\eta \cos \theta, \eta \sin \theta, \sqrt{1-\eta^2}),
\label{eq:magBEC}
\ee
where $\eta=\sqrt{2\tilde \eta/(1+\tilde \eta^2)}$ is a measure of the local magnon-condensate fraction
$\tilde \eta$ \cite{Giamarchi2008}, and $\theta$ is
the local in-plane angle. That this magnetization configuration indeed corresponds
to a magnon condensate is understood by noting that $n_x \sim \mathrm{Re} \langle \hat b \rangle$
and $n_y \sim \mathrm{Im} \langle \hat b \rangle$, where $\hat b$ is the Holstein-Primakoff operator
that annihilates a magnon.

Using Eq.~\eqref{eq:magBEC} together with the Landau-Lifshitz equation Eq.~\eqref{eq:lle}, and using
the mean-field equations for the total density, we obtain a set of hydrodynamic equations, governing the evolution
of the magnon condensate and the scalar condensate at zero temperature:
\ba
D_t \rho/\rho = -\bi\nabla \cdot \bi v,
\,\,\,
M D_t \bi v 
= -\bi \nabla \mu,\\
D_t n_z
+ \bi v_\mathrm{spin} \cdot \bi \nabla \rho / \rho
=-\bi \nabla \cdot \bi v_\mathrm{spin},
\,\,\,
\hbar D_t \theta
=
-\mu_\mathrm{spin}
,
\label{eq:MagnonEoms}
\ea
where
$
\mu =
(g_0+g_1) \rho 
+ \hbar^2 (\bi \nabla n_\mu)^2/4M
$,
and
$
\mu_\mathrm{spin}/\hbar J =
n_z (\bi \nabla \theta)^2
-(\bi \nabla^2 \eta)/n_z \eta
-
(\bi \nabla \eta)^2/n_z^3
-Kn_z/J
%
$
are the chemical potentials. Their gradients drive the mass and spin
supercurrents.
Note that the quantum pressure term \cite{PethickSmithBook}
has been neglected from the scalar condensate equations, as we
will only be interested in states of uniform particle density
from hereon.
Therefore, these hydrodynamic equations only describe density
dynamics on length scales longer than the condensate healing
length.
On the other hand, similar terms have been kept for now in $\mu_\mathrm{spin}$
since we consider an inhomogeneous magnon condensate in Sec.~\ref{ss:close}.
Furthermore, even though similar equations have been derived in previous work
\cite{Stamper-Kurn2013,UedaHydro,Lamacraft2008,Oh2014,Ueda2012},
here we describe spin and mass superfluidity in a single framework,
which has not been done before to the best of our knowledge.
Hence, these hydrodynamic equations constitute the central 
result of our article.

The total velocity
can be separated into two parts \cite{UedaHydro}, that can be addressed separately \cite{Madison2000,Choi2013,CoherentMagnonOptics}. First, we have the conventional
superfluid velocity $\bi v_\mathrm{scalar} = (\hbar/M) \bi \nabla \phi$
due to the phase $\phi$ texture of the atomic condensate wavefunction.
However, there also is the spin superfluid velocity $\bi v_\mathrm{spin} = -J\eta^2\bi \nabla \theta$ due to the phase $\theta$
texture of the magnon condensate.
Thus, the full velocity in this ferromagnetic spinor Bose condensate is 
\be
\bi v = \bi v_\mathrm{scalar} + 2n_z \bi v_\mathrm{spin} /\eta^2.
\label{eq:fullvelocity}
\ee
This velocity $\bi v$ influences the magnetization dynamics through the
advection term in the Eulerian derivative 
since the magnetic moments (individual atoms) are mobile in a cold-atom 
system, similarly to e.g.\ so-called ferromagnetic superconductors \cite{Pfleiderer2009}. This is in contrast to most solid-state ferromagnets, where the magnetic moments are
localized.

\subsection{Linear regime and collective modes}
In the linear regime, the two superfluids are decoupled as the Eulerian derivatives become ordinary derivatives. 
The elementary excitations
on top of the homogeneous magnon condensate ($\eta, \theta=
\mathrm{const.}$) have a dispersion
\be
\omega^2 = K J \eta^2 k^2 + J^2 k^4,
\label{eq:dispersion}
\ee
which follows from the equations of motion above, and is linear
in the long-wavelength limit \cite{Halperin1969}. Moreover, the dispersion
of the density excitations follow the Bogoliubov dispersion \cite{Stamper-Kurn2013},
\be
\omega^2 = (\hbar/2M) k^2 [ (\hbar/2M) k^2 + 2 (g_0+g_1) \rho/\hbar ],
\ee
the derivation of which requires keeping the quantum
pressure terms (see Ref.~\cite{UedaHydro} for an explicit calculation).
At long wavelengths the dispersion is linear in both cases, 
with the speed of sound 
equal to $c_\mathrm{spin}=\sqrt{K J \eta^2}$ for the spin excitations
and $c_\mathrm{scalar} = \sqrt{(g_0+g_1)\rho/M}$ for the density
excitations, signaling that we are dealing with a superfluid,
as using the Landau argument \cite{Landau1941} one can show that excitations travelling with velocities
slower than the lower of these two speeds of sound are not damped.
Note that in general spinor gases are not necessarily spin superfluid in the
ferromagnetic regime. In particular, the spin currents discussed 
in Ref.\ \cite{KU02} in the absence of quadratic Zeeman effect
are not spin supercurrents, as their critical velocity vanishes.
From this point on we only consider the critical spin superfluid velocity
and drop its subscript: $c \equiv c_\mathrm{spin}$.
Note that this homogeneous state is stable and displays spin 
superfluidity in the easy-plane situation $K>0$ only.  
Furthermore, note that tilting the magnetization,
i.e., considering the magnon condensate in the ferromagnetic
phase, results in a spectrum that is distinct from both
the usual collective modes in the ferromagnetic phase
as well as the broken axisymmetry phase (c.f.~the appendices
of Ref.~\cite{Murata2007}).

In order to verify that the
system is indeed a spin superfluid and to show that the conventional
superfluidity and the spin superfluidity stand on an equal footing, we propose two experiments which should be
realizable with current experimental techniques. To keep the description
as simple as possible, we work in the one-dimensional limit, where the spatial
confinement is strong in two dimensions and more gentle in the remaining 
spatial dimension. Furthermore, in order to avoid complications due to trap averaging, 
we consider a box trap \cite{Gaunt2013}. 
A boxlike potential of a form similar to the one depicted in Fig.~\ref{fig:exp0}
has already been experimentally produced and persistent currents of a scalar superfluid 
have been observed in such a setup in Ref.~\cite{Corman14}.

\subsection{Far-from-equilibrium spin superfluidity signature in the nonlinear regime}
\label{ss:far}
One of the hallmarks of superfluidity
is an unobstructed flow of current. In particular, in a spin superfluid, a spin current can
flow with no dissipation as opposed to a system with diffusive spin transport \cite{Niroomand2015}, where
the spin current decays after traversing some finite length, which depends on the
diffusion length and on the timescales of various spin relaxation mechanisms \cite{Sonin2010}. 
In this section we describe a stationary
state (Fig. \ref{fig:exp0}) where 
the supercurrent and the spin supercurrent flow in opposite
directions, resulting in no mass transport. However, since the spin
current flows through the whole sample, it thus illustrates dissipationless
spin transport \cite{Sonin2010}.

By considering linear gradients in the atomic condensate phase $\phi$ and in the magnon condensate phase $\theta$
with a constant magnon condensate fraction $\eta$, from Eqs. 
\eqref{eq:MagnonEoms} we find
the stationary-state condition
\be
\phi' = \frac{n_z}{2}\left( \theta' + \frac{K}{J \theta'} \right),
\ee
where the primes indicate spatial derivatives. In order to show that
this condition can be satisfied for realistic experimental 
parameters, we consider a concrete example of a ring of length $L$
filled with a cloud of spin-1 atoms (e.g.\ the $F=1$ hyperfine manifold of
the strongly ferromagnetic $^{7}$Li) 
prepared in the magnon-condensate state described by
Eq.~\eqref{eq:magBEC}
In that case, a state with $n_z=1/2$ where the atomic condensate phase
winds once ($\phi' = 2\pi/L$), while the magnon condensate phase has two windings ($\theta'=2\times 2\pi/L$),
is stationary for ${KL^2/\pi^2J}=16$. For a ring of length $L =
50\mu\mathrm{m}$, we have $J\pi^2/L^2 \simeq 20$Hz, which requires a moderate
easy-plane anisotropy of circa $300$Hz.
As explained in Sec.~\ref{sec:framework}, in order for our theory 
to be applicable, the gap $\Delta_\mathrm{FM}=g_1 \rho$ has be larger than any
other energy scale relevant to the dynamics.
This requirement can be satisfied
for a strongly ferromagnetic gas of a sufficiently high density. In the case of 
the stable $\mathrm{F}=1$ manifold of the $^{7}\mathrm{Li}$ atom \cite{Stamper-Kurn2013},
a relatively high but experimentally feasible density of
$10^{20}\mathrm{m}^{-3}$ leads to a gap of $\Delta_\mathrm{FM}/\hbar = 3.5 \mathrm{kHz}$
which is well above any other energy scales relevant to the dynamics considered here.
Moreover, this static state implies no mass transport as $\bi v_\mathrm{scalar}$
exactly cancels $\bi v_\mathrm{spin}$,
and thus $\bi v=0$. 
Furthermore, this stationary state is also stable according to the
relevant Landau criteria \cite{Landau1941}. When it comes to the spin supercurrent,
$|\bi v_\textrm{spin}|/c_\textrm{spin} = 1/2 < 1$.
Concerning density excitations, the criterion is satisfied due to the
large gap $\Delta_\mathrm{FM}$ for the densities in question, namely,
\be
|\bi v_\textrm{scalar}|/c_\textrm{scalar}
=
\sqrt{
8 \frac{J \pi^2 / L^2}{(g_0+g_1)\rho/\hbar}
}
\ll 1.
\ee
Note that without easy-plane anisotropy, and therefore without spin superfluidity, such
steady states cannot be obtained.

In particular, we would like to emphasize the difference between the state described
in this subsection and the fractional vortices in polar spinor condensates.
When it comes to the polar (nonmagnetic) phase, the dispersion of spin
amplitude or the nematic angle is indeed linear at long wavelengths, signaling
a nonzero critical (Landau) velocity for these excitations \cite{UedaHydro}. 
However, the dynamics of
this system is governed by the nematic director, in sharp contrast to
conventional magnetism governed by the magnetization vector dynamics.
More formally, the order-parameter space in the polar
phase is $\mathrm U(1) \times \mathrm S^2 / \mathbb Z_2$. 
This structure comes from the fact that a $\pi$ rotation
of the scalar phase in combination with inverting the nematic director
leaves a polar state unchanged. Indeed, this symmetry in particular allows 
fractional vortices to exist in the polar state. 
Therefore, even though the current defined in terms of the gradients
of the nematic director is sometimes called the spin supercurrent\cite{KU02}, 
it is in our opinion to be
distinguished from the usual case, i.e., from the spin current defined 
in terms of a gradient of magnetization direction, as is conventional
in other systems, such as solid state magnets.
Finally, in the ferromagnetic phase of spin-1 bosons that is considered in this
article, the order parameter is
indeed the magnetization vector, and its space is the usual SO(3).  In that
case the mass current has a spin-texture contribution 
(c.f.~Eq.~\eqref{eq:fullvelocity}),
which allows us to find the stationary states described in our work. 
In the special case of easy-plane anisotropy the spin current carried by the texture is a supercurrent.
No such
spin-texture term exists in the polar phase\cite{KU02}, 
and thus the physical mechanism behind the half-quantum vortices is
in our opinion remarkably different from the stationary states that we propose.

\subsection{Close-to-equilibrium spin superfluidity signature}
\label{ss:close}
\begin{figure}[t]
\begin{center}
\includegraphics[width=0.49\linewidth]{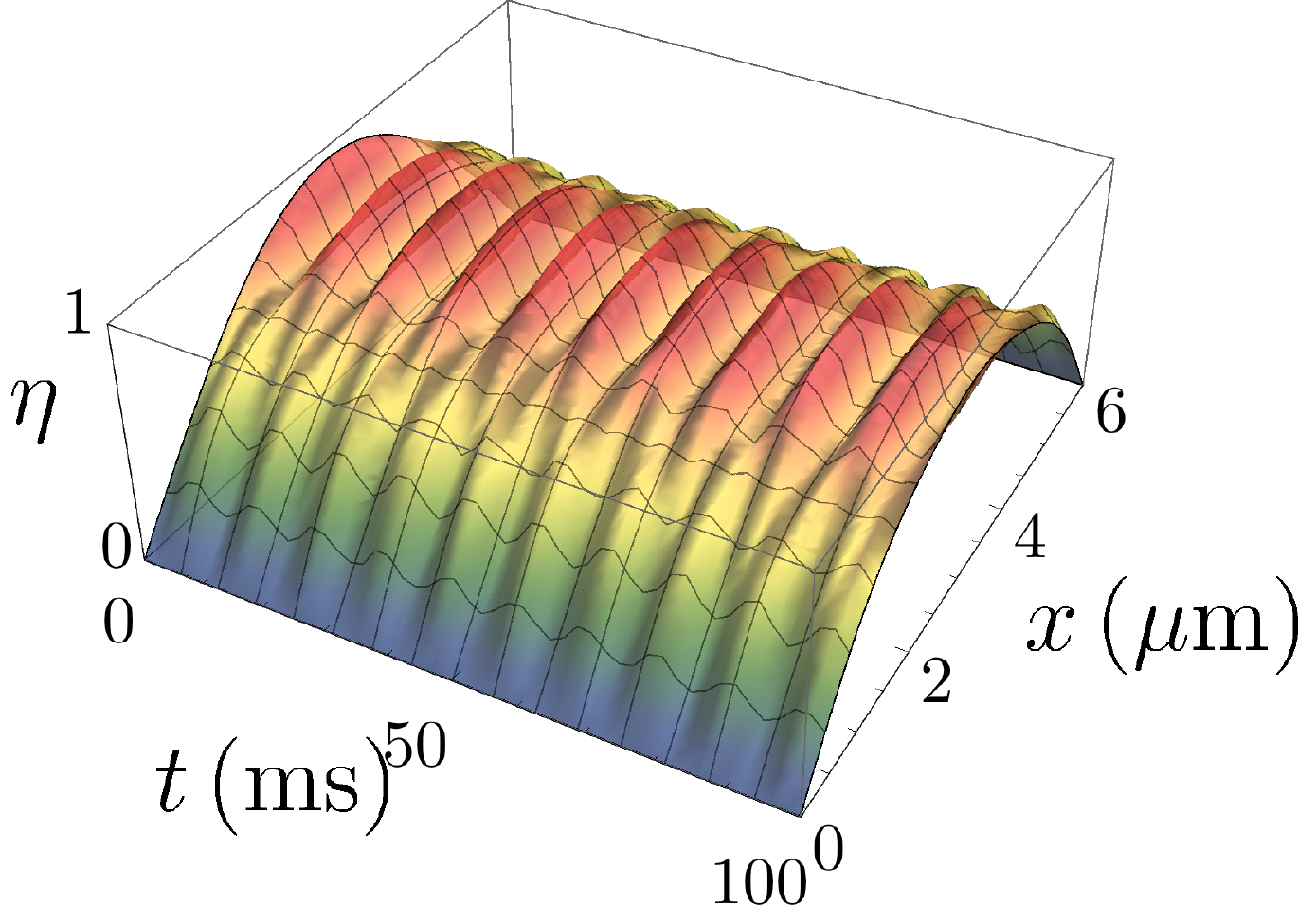}
\includegraphics[width=0.49\linewidth]{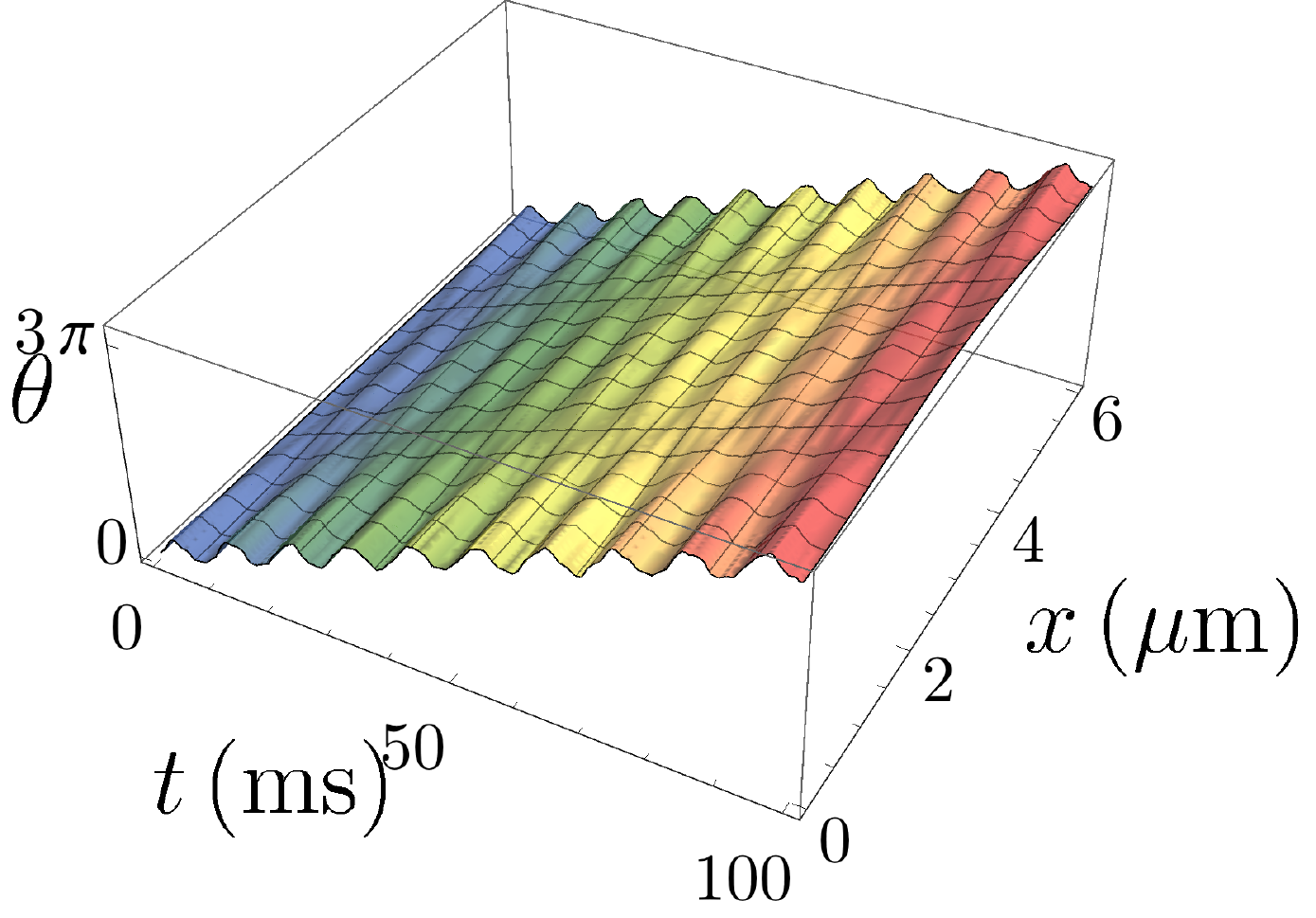}
\caption{%
Evolution of a sine-shaped magnon condensate packet for realistic experimental parameters:
an $L=20\mu\mathrm{m}$-long cloud of strongly ferromagnetic 
$^{7}$Li atoms, with the resulting exchange 
$J\pi^2/L^2 \simeq 100$Hz, easy-plane anisotropy of $K=300$Hz, and a peak initial
condensate density of $\eta_0=0.95$ according to
Eqs.~\eqref{eq:MagnonEoms} with no approximations.
The density of the
condensate $\eta$ (left) is almost unaffected by the dynamics, while the in-plane angle
$\theta$ (right) evolves almost uniformly (see Sec.~\ref{ss:close} for more details). The in-plane
magnetization completes a full rotation in approximately $60$ms. Note that the
total magnetization in the $z$ direction is conserved during the evolution.
}
\label{fig:exp1}
\end{center}
\end{figure}

It is also possible to observe a signature of spin superfluidity by measuring the
time evolution of the in-plane magnetization angle $\theta$ in the 
simple ``bar'' geometry (as opposed to the ring discussed above). 
To that end, consider a magnon condensate
with a constant atomic condensate phase $\phi$ and a constant in-plane angle $\theta$, in addition 
to a smoothly varying
magnon condensate fraction bump $\eta = \eta_0 \sin(\pi x/L)$ at the initial time $t=0$, where $L$ is the
length of the atomic cloud and $x$ is the spatial coordinate, 
c.f.~Fig.~\ref{fig:exp1}. In this case, up to the
lowest order in the gradient expansion,
the time evolution preserves the magnon-condensate-density profile such that $\partial_t \eta=0$,
while the in-plane angle rotates in time with no spatial profile developing,
\be
\partial_t \theta = K\langle n_z \rangle - J\pi^2/\langle n_z \rangle L^2,
\ee
where
$
\langle n_z \rangle =2{\mathcal E}[\eta_0^2]/\pi
$
is the $z$ component of magnetization averaged over the length of the cloud, and 
${\mathcal E}$ is the complete elliptic integral.

We now connect this inhomogeneous setup with the magnon dispersion 
of the homogeneous
system in Eq.~\eqref{eq:dispersion}. On the one hand, local-density
approximation \cite{PitaevskiiStringari2016} is applicable only when 
the relevant quantity ($\eta$ in this case) varies slowly enough with 
respect to the appropriate coherence length (also known as the healing length) 
$\xi$ \cite{PethickSmithBook}. On the other hand, a system with
a linear dispersion can only be considered superfluid, if its evolution
actually probes the linear part of the dispersion. 
Here the healing length
\be
\xi = \sqrt{\frac{J}{K}}\frac{1}{\eta} 
\label{eq:coherenceLength}
\ee
is obtained by comparing the two terms in Eq.~\eqref{eq:dispersion}.
The relevant length scale to compare with $\xi$ is in this case
the length of the atomic cloud $L$. 
In the center of the cloud, where $\eta \simeq 1$, using the parameters
in the caption of Fig.~\ref{fig:exp1}, we have $\xi / L \simeq 0.2 \ll 1$.
Therefore, we conclude that the local-density approximation is applicable in 
the central part of the cloud.
We now check if the linear part of the dispersion is actually probed.
Since the longest possible wavelength 
is equal to the system size, $L$, we plug
the wavevector $k=\pi/L$ into the dispersion relation in Eq.~\eqref{eq:dispersion} and obtain 
$(\pi/L)^{-2} \omega^2 = K J \eta^2 + J^2 (\pi/L)^2$.
Thus, to see if the linear part of the dispersion is relevant, we have
to compute the ratio between two sound velocities, namely, $c^2=K J \eta^2$ and $c_0^2=J^2 (\pi/L)^2$. 
Furthermore, since $c^2$ is position dependent, we
have to average it: $\langle c^2 \rangle = K J \eta_0^2/2$.
It can be concluded that spin superfluidity becomes pronounced when
\be
\frac{\langle c^2 \rangle}{c_0^2}=\frac{K J \langle \eta^2 \rangle}{J^2 (\pi/L)^2} 
= \frac{K \eta_0^2}{2J (\pi/L)^2} > 1.
\ee
This criterion can be readily evaluated experimentally by observing the evolution
of the in-plane angle. In particular, the evolution stops ($\partial_t \theta = 0$)
when 
$
KL^2 \langle n_z \rangle^2
/
J\pi^2
=1
$.
Therefore, prevalence of spin superfluidity in the sample
for large condensate fractions ($\eta_0 > 0.95$)
is proven by measuring that $\partial_t \theta > 0$,
since ${\mathcal E}[\eta_0^2]$ is a monotonically decreasing function,
and since $\langle n_z \rangle^2(\eta_0\simeq 0.95) \simeq 1/2$.

Hence, for a large condensate fraction $\eta$ observing the evolution of
$\theta$ is sufficient to distinguish between the situation where the critical velocity $c$ vanishes,
and where it is substantial. Namely, if $\partial_t \theta < 0$, the system is dominated
by quadratic excitations and has a negligible critical velocity, whereas if
$\partial_t \theta > 0$, the system displays spin superfluidity. In order to
illustrate that our conclusion holds for the full solution
for realistic experimental parameters, we plot the numerical solution 
of Eqns.~\eqref{eq:MagnonEoms} with the boundary conditions
$\eta(0)=\eta(L)=0$ and $\theta'(0)=\theta'(L)=0$ and the aforementioned
initial conditions in Fig.~\ref{fig:exp1}.

\subsection{Incoherent dynamics}
\label{ss:incoherent}
Up to this point we have only considered 
coherent magnetization dynamics. However, due to nonzero temperatures and 
interparticle collisions present in real ultracold-atom systems, it is 
important to investigate incoherent (kinetic) processes as well.
In particular, it is interesting to study the time scales of the incoherent 
(kinetic) magnon dynamics and compare them to the time scales of the coherent 
evolution. We compare these two timescales by
evaluating the dominant incoherent timescale set by 
the four-magnon interaction \cite{Dyson1956,Keffer1961,Harris1968,Halperin1969} kinetic integral,
which we obtain by using the Holstein-Primakoff transformation \cite{Holstein1940}
in order to describe the spin degrees of freedom in this ultracold-atom system in terms of magnons.
In particular, we compute that the
coherent dynamics discussed previously is an order of magnitude faster (tens of milliseconds)
than the incoherent processes (hundreds of milliseconds)
for an experimentally accessible system 
(see~Refs.~\cite{CoherentMagnonOptics,ThermometryDemagCooling}).
This implies that, while our hydrodynamic description is valid at zero temperature, at finite 
temperatures the thermal magnons do not equilibrate on the timescales set by the coherent 
dynamics, which in turn means a hydrodynamic description of the spin dynamics cannot be readily 
obtained at finite temperature. 

In order to compute the four-magnon scattering time,
we consider the exchange term in the spinor Bose gas Hamiltonian, namely,
$J :\boldsymbol {\hat\Omega} \cdot \bi \nabla^2 \hat{\boldsymbol \Omega}:$,
where $\boldsymbol {\hat \Omega}$ is the full magnetization operator.
It can be divided into the magnetization density $\rho_s$,
which we assume to be constant in the deep ferromagnetic regime, and
the direction of magnetization operator $\boldsymbol {\hat n}$ in the following
way: $\boldsymbol {\hat\Omega} = \rho_s \boldsymbol {\hat n}$.
Hence, the
magnetic excitations only concern the direction of magnetization
$\bi n$ in this regime. 
Since we wish to merely compare the order of magnitude of the coherent
and incoherent dynamics, we do not consider the quadratic Zeeman effect
in this calculation. As discussed below, making reliable quantitative
predictions requires not only including the quadratic Zeeman effect,
but also taking into account magnetic field inhomogeneity.
The Holstein-Primakoff transformation
\cite{Holstein1940,auerbachbook} introduces
bosonic magnon operators $\hat b(\bi k)$, which substitute for
the magnetization direction operators. We subsequently perform
a semiclassical expansion of these spin fluctuations around the average
direction of the magnetization,
and retain only two first terms in this large-magnetization expansion.

The lowest-order
term yields the kinetic energy for the magnons,
\be
\mathcal H_2 = \int \frac{d\bi k}{(2\pi)^3} E_k \hat b^\dagger(\bi k) \hat b(\bi k),
\ee
where $E_k = \hbar J\bi k^2$ is the magnon dispersion
\cite{KranendonkVleck1958,Akhiezer1959},
whereas the subleading term describes the four-magnon interaction,
\ba
\mathcal H_4 &=&
\int
\frac{\dd \bvk}{(2\pi)^3}
\frac{\dd \bvk_2}{(2\pi)^3}
\frac{\dd \bvk_3 }{(2\pi)^3}
\frac{\dd \bvk_4}{(2\pi)^3}
\delta^{(3)}(\bi k + \bi k_2 -\bi k_3 - \bi k_4)
\nonumber \\
&\times&
g\,
\hat b^\dagger (\bi k)
\hat b^\dagger (\bi k_2)
\hat b(\bi k_3)
\hat b(\bi k_4),
\ea
where
\be
g=\frac{\hbar J}{4\rho_s}
\left(
\bi k \cdot \bi k_2
+
\bi k_3 \cdot \bi k_4
\right)
\ee
is the four-magnon coupling constant.
Note that this coupling constant $g$ does not explicitly depend on the scattering properties
of the particular atom.
We also note that including the easy-plane anisotropy resulting from the quadratic Zeeman shift
leads to a constant (momentum independent) contribution to the scattering amplitude that we
ignore here.
From this interaction term, using e.g.\ the Fermi Golden Rule,
we construct a collision integral. The characteristic timescale is thus given by
\ba
\frac{1}{\tau_{\bi k}}
&=&
\frac{2\pi}{\hbar}
\int \frac{d \bi k_2}{(2\pi)^3}
\int \frac{d \bi k_3}{(2\pi)^3}
\int \frac{d \bi k_4}{(2\pi)^3}
\, g^2 \, \mathcal F
\\ \nonumber
\times
\,
\delta(E_{\bi k} &+& E_{\bi k_2} - E_{\bi k_3} - E_{\bi k_4})
\,
(2\pi)^3 \delta^{(3)}(\bi k + \bi k_2 - \bi k_3 - \bi k_4)
,
\label{eq:1tau}
\ea
where
\be
\mathcal F
=
f_2 (1+f_3) (1+f_4)
-
(1+f_2) f_3 f_4
\ee
is a combination of Bose-Einstein distributions
$f_i = 1/[\exp(\beta E_{\bi k_i})-1]$,
and $\beta = 1/k_BT$ is the inverse thermal energy.
\begin{figure}[t]
\begin{center}
\includegraphics[width=\linewidth]{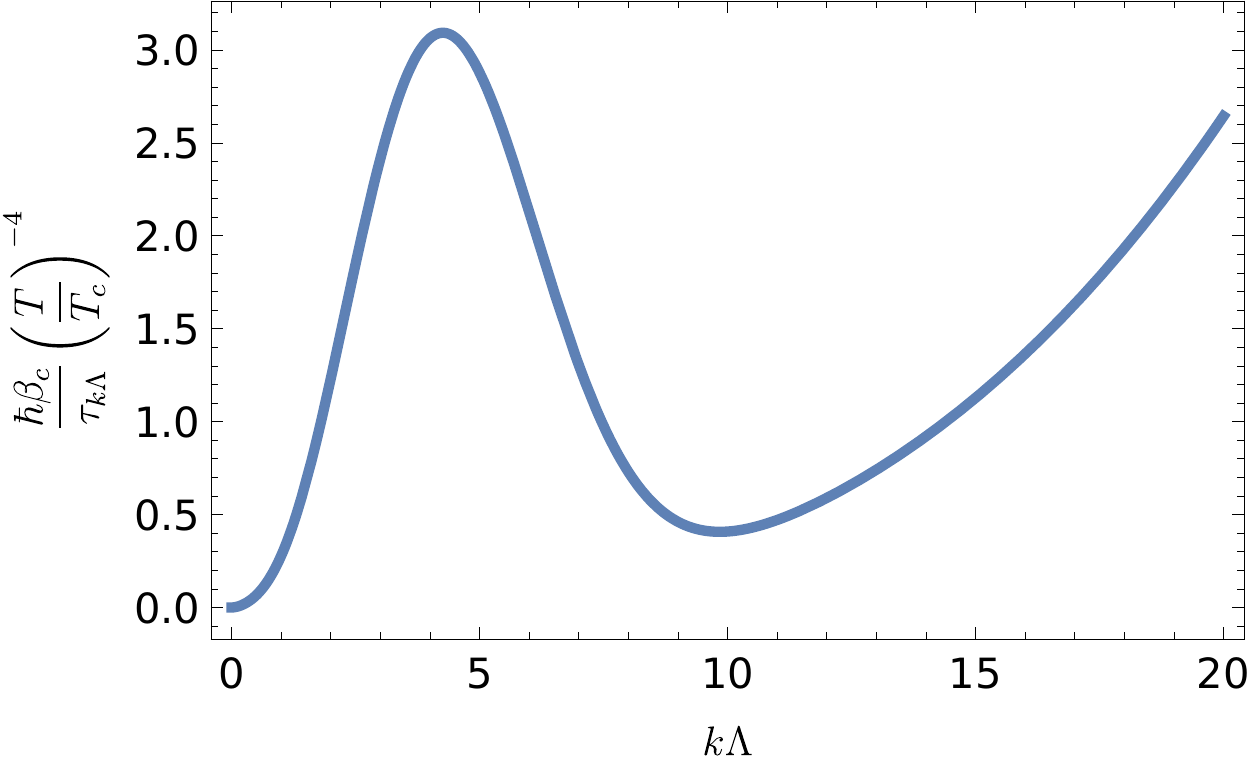}
\caption{
Typical incoherent dynamics timescale evaluated from the four-magnon interaction term.
}
\label{fig:kinetic}
\end{center}
\end{figure}
This timescale can be written in the dimensionless form,
\be
\frac{\hbar\beta_c}{\tau_{k\Lambda}} = \left(\frac{T}{T_c}\right)^4 I(k\Lambda),
\ee
where $\Lambda$ is the thermal magnon de Broglie wavelength,
$k_BT_c \simeq 4\pi\hbar J (\rho_m/2.6)^{2/3}$ is the condensation
temperature of a noninteracting homogeneous magnon gas with density $\rho_m$,
and $I(k\Lambda)$ is a dimensionless function, which we evaluate numerically
(see Fig.~\ref{fig:kinetic}).

Note that for
small $k\Lambda$, the kinetic frequency $\tau_{k\Lambda}^{-1}$ approaches zero; as $k\Lambda$ 
increases,
$\tau_{k\Lambda}^{-1}$ has a local maximum and a local minimum, and only then does it
approach the classical limit.
These features can be understood in terms of competition between the Bose enhancement
and the four-magnon coupling constant, as the former favors low $k\Lambda$ scattering, while 
the latter suppresses it.
At low momenta, fitting $I$ yields a function
$(k\Lambda)^2\log k\Lambda$, similar to the one reported previously in a field-theoretical calculation \cite{Harris1968,Halperin1969}.

Before comparing the result of our kinetic theory
with experimental findings, we
note that our result is the lower bound on the relaxation frequency,
since we have only included a single relaxation process.
It is likely that due to other processes (e.g. related to
magnetic-field gradients), the relaxation is faster. This statement
is consistent with the fact that for a realistic system
our result is $1/\tau \simeq 0.2$Hz, i.e., an order of magnitude lower than the recently reported experimental rate \cite{CoherentMagnonOptics}.
The comparison is performed using the following data.
From Fig.\ 1 (b) in the paper on coherent magnon optics (Ref.\ \cite{CoherentMagnonOptics}) we
read off that $1/\tau \simeq 4$Hz for $k=2\pi/(15.4\mu\mathrm{m})$. No temperature is given in the paper, but
assuming that $T=T_c/10$, and taking $T_c \simeq 1\mu K$ from another paper of the same group
(Ref.\ \cite{ThermometryDemagCooling}), we have $k\Lambda\simeq1/4$ and $\hbar\beta_c \simeq 10^{-5} s$.
These assumptions yield $1/\tau \simeq 0.2$Hz, which is one magnitude lower than the experimental result \cite{CoherentMagnonOptics}.
Note that it is possible to improve our calculation by performing an accurate trap average, distinguishing
between magnon and gas condensation temperatures and populations etc.

As discussed above, while our hydrodynamic description is valid at zero temperature, at finite 
temperatures the thermal magnons do not equilibrate on the time scales set by the coherent 
dynamics, which in turn means a hydrodynamic description of the spin dynamics cannot be readily 
obtained at finite temperature.
This stands in contrast to the
situation regarding the scalar degrees of freedom \cite{Miesner1998} in this system.
The difference is mainly due to the different coupling constants, as
the four-magnon coupling constant depends on the density
and momenta of the magnons, whereas for the scalar degrees of freedom the coupling constant 
only depends on the $s$-wave scattering length $a$ of the atom \cite{Leggett2001}.
A typical collision time of atoms in the classical approximation \cite{Meppelink2009} is 
$\tau_\mathrm{cl} = 1/n \sigma v_\mathrm{rel}$, where $n$ is the density, $\sigma=8\pi a^2$ is 
the collision cross section,
and $v_\mathrm{rel}$ is the relative thermal velocity.
For a homogeneous $^{7}$Li gas with a density of $10^{20}\mathrm{m}^{-3}$ considered
in Secs.~\ref{ss:far}~and~\ref{ss:close}, the condensation temperature is $5\mu$K. 
At one tenth of this temperature, the collision time of atoms is $2$ms%
\footnote{
For a homogeneous $^{87}\mathrm{Rb}$ gas comparable to the experiments in
Refs.~\cite{CoherentMagnonOptics,ThermometryDemagCooling} with a density
corresponding to the condensation temperature 
of $1\mu$K, at the temperature of $0.1\mu$K, the collision time of atoms is $0.6$ms.
}%
. Therefore, at a nonzero temperature such a system is in the hydrodynamic regime
regarding the scalar degrees of freedom, which is enforced by
rapid collisions. On the other hand, given the same conditions the magnon dynamics is in the
collisionless regime, allowing for far-from-equilibrium states to persist for long periods of 
time.
We reiterate that this timescale hierarchy is well defined, as different timescales are 
separated by at least one order of magnitude.
Finally, the above-described situation concerning magnons in a spinor gas is also different from
the binary-mixture situation. There, longitudinal spin kinetics is comparable to the
incoherent dynamics of the scalar fields, and hence longitudinal spin currents relax in milliseconds as well \cite{Armaitis2015,Koller2015}.

\subsection{Lower and upper spin currents}

In a general spin superfluid, in addition to an easy-plane anisotropy $K$, an
$n_x$-fold in-plane anisotropy $K_x$ can play an important role \cite{Sonin2010,Duine2015}.
This in-plane anisotropy favors the magnetization to point along one of the $n_x$ axes in the easy plane.
The interplay
between the exchange energy and each of these two different
anisotropies defines the upper critical spin current $j_{c,up}$ and the lower critical
current $j_{c,low}$, respectively.

In particular, when the in-plane spin rotation $|\boldsymbol \nabla \theta|$ is so large, 
that the corresponding spin current $j$ exceeds $j_{c,up}$,
this spin texture can relax by escaping the easy-plane, i.e.\ acquiring a magnetization
component along the hard axis.
The upper critical spin current can thus be estimated by equating
the exchange energy and the energy of the easy-plane anisotropy:
\be
J|\boldsymbol \nabla \theta|^2_{c,up} = \hbar K.
\ee
On the other hand,
if the in-plane spin rotation $|\boldsymbol \nabla \theta|$ is sufficiently small
($j < j_{c,low}$), a single spin domain wall arises, thus
preventing superfluid spin flow over macroscopic distances. Hence, the lower critical spin current can be derived by equating the exchange energy and the energy corresponding to the in-plane anisotropy:
\be
J|\boldsymbol \nabla \theta|^2_{c,low} = n_x \hbar K_x.
\ee
Therefore, for a large range of
spin currents to be allowed, it is required that $j_{c,low} \ll j_{c,up}$, or, equivalently,
the in-plane anisotropy has to be much weaker than the easy-plane anisotropy:
$K_x \ll K$.

When it comes to cold-atom systems, the easy-plane anisotropy $K$ is well-controlled
and can be made large compared to the exchange energy, as described in the main text.
We expect that the in-plane anisotropy, however, will be chiefly
caused by stray magnetic fields.
It thus will presumably be small as compared to the easy-plane anisotropy, and already
controlled for in the experiment for other reasons. We note
further that magnetic dipole-dipole interactions can also give rise to a lower critical
spin current. However, for single atoms these magnetic dipole-dipole interactions are usually
weak, and can be minimized by choosing a suitable atomic species, see Ref.~\cite{Lahaye2009}
for a review.
For the two states proposed in the article, the ratios between the exchange energy and the
easy-plane anisotropy are $0.06$ (see Sec.~\ref{ss:far}) and $1/3$ (see the caption of Fig.~\ref{fig:exp1}). We are hence confident
that both proposed states correspond to spin currents, which belong
to the region between the lower and the upper critical spin currents.

\section{Summary and future work}
In summary, we have examined the zero-temperature coherent dynamics of spinor Bose gases
 with a quadratic Zeeman effect and shown that this system exhibits coexisting spin and mass superfluidity.
We have described
two experimentally-accessible states illustrating the interplay between superfluidity and spin superfluidity.
By evaluating a four-magnon collision integral and 
comparing the relevant timescales, we have concluded that at nonzero temperatures the magnon gas is
in the collisionless regime
in stark contrast to the situation concerning the scalar degrees of freedom. 

Experimental access to the states described in this work is constrained by the 
condition that the spin-dependent interaction has to be much larger
than all the other relevant energy scales. At present, it seems that this
constraint can be most readily satisfied using strongly ferromagnetic
atomic species, such as $^7$Li. However, other ways in addressing 
this restriction, such as employing Feshbach resonances \cite{Chin2010}
to enhance interactions, or investigating systems with high particle
density, e.g., the $^3$He liquid \cite{Vollhardt2013} 
or solid-state superconductors with
spin order \cite{Linder2015} might be possible.

In the future work, we plan to extend our description by including the experimentally-relevant
magnetic-field inhomogeneities into our LL equation, which will allow us to make a direct
connection to the recent experimental results on magnon condensation
\cite{Fang2015,ThermometryDemagCooling}. Another interesting direction is to consider
nonzero temperature dynamics of scalar and spin degrees of freedom, which is
particularly intriguing in this system due to a clear hierarchy of timescales involved. Finally,
we point out that coupling between magnon condensate and thermal cloud is facilitated by the quadratic
Zeeman term \cite{Flebus2016} which could make the experimental control over the build-up of
magnon coherence and condensate growth possible. Moreover, this anisotropy will also affect the magnon kinetics.

\section*{Acknowledgements}

It is our pleasure to thank Henk Stoof and Gediminas Juzeli\={u}nas for stimulating discussions.
J.~A. has received funding from the European Union's Horizon 2020 research and innovation
programme under the Marie Sk\l odowska-Curie grant agreement No 706839 (SPINSOCS).
R.~A.~D. is supported by the Stichting voor Fundamenteel Onderzoek der
Materie (FOM) and is part of the D-ITP
consortium, a program of the Netherlands Organisation for Scientific Research (NWO)
that is funded by the Dutch Ministry of Education, Culture and Science (OCW).

\bibliography{main}

\end{document}